# Experimental and Computational Demonstration of a Highly Stable, in-situ Pt Decorated Sputtered ZnO Hydrogen Sensor for sub-ppm Level Detection

*Puja Ghosh[1], Pritam Ghosh[2], Rizwin Khanam[1], Chandra Shekhar Prajapati[3], Aarti Nagarajan[4], Shreeja Das[5], Rakesh Paleja[6], Sharan Shetty[5], Gopalakrishnan Sai Gautam[2], Navakanta Bhat[1, \*]*

[1]Centre for Nanoscience and Engineering, Indian Institute of Science, Bengaluru, [2]Department of Materials Engineering, Indian Institute of Science, Bengaluru, [3]Indian Institute of Technology Patna, [4]Shell Global Solutions (US) Inc., [5]Shell India Markets Pvt. Ltd., [6]Shell Information Technology International Ltd.

*Corresponding author. E-mail address: navakant@iisc.ac.in; navakant@gmail.com

**Abstract**

In this work, we present a Pt decorated ZnO thin film-based gas sensor for hydrogen detection, fabricated using a sputtering technique and an in-situ Pt decoration approach. Our design yields a stable, highly sensitive, and repeatable response to hydrogen gas. Our sensor demonstrates optimal performance at an operating temperature of 225 °C, with rapid response and recovery times (~10 sec and 3 sec), high selectivity, and long-term stability. Specifically, we deposit a ZnO thin film on an interdigitated electrode (IDE) substrate, with Pt nanoclusters added to the (002) polar plane by brief sputtering (1 to 6 sec) to create an active sensing interface. We find the Pt decorated ZnO sensor, with a Pt deposition time of 2 sec, to exhibit an enhanced response (~52,987%) to 1% hydrogen concentration, indicating its suitability for industrial and environmental monitoring applications. Additionally, our device demonstrates reliable detection of low hydrogen concentrations (~100 ppb), with a response of ~38% and no response drift over one year of testing, underscoring the long-term stability of the sensor. To elucidate the role of Pt deposition and pristine ZnO in hydrogen sensing, we performed density functional theory calculations, analysing adsorption and reaction energetics involving adsorbed $H_2$, $O_2$, O, OH, and $H_2O$, and lattice oxygen atoms on the ZnO (002) surface with and without Pt decoration. Our computational data is in agreement with our experimental observations, identifying the oxygen-exposed (002) surface to be most active for hydrogen sensing in both pristine and Pt decorated ZnO. Further, our computations highlight the role of Pt in enhancing hydrogen sensitivity via *i*) activating an autoreduction pathway of adsorbed hydroxide species, *ii*) spontaneous dissociation of adsorbed molecular hydrogen, and *iii*) keeping the lattice oxygen pathway of forming water active. Our systematic approach of designing sensors combining a robust experimental setup with theoretical insights, are key in developing efficient hydrogen gas sensors, as well as in understanding the mechanisms behind such superior performance.

## 1. Introduction

The global interest to transition towards cleaner energy sources has positioned hydrogen as a pivotal component in the transition to sustainable energy systems.[1] Due to its high flammability in air that is indicated by the low explosive limit (LEL), in volumetric terms, of only 4%, reliable hydrogen monitoring systems are essential to ensure safety and detect any potential leaks. While a comprehensive discussion on the current overview of hydrogen sensing technologies and the evolving requirements of sensor performance for hydrogen systems can be found in literature[2,3], current research efforts are focused on bridging the gap in designing fast, highly sensitive, and stable hydrogen sensing materials. Apart from its use as a renewable fuel, recent studies suggest that hydrogen has an indirect global warming potential over a 100-year period, estimated to be ~12.[4] Hence, it becomes important not only to have hydrogen sensors for safety but also for highly sensitive and accurate monitoring of hydrogen levels in the atmosphere for environmental applications. Thus, it is important to develop sensors that can reliably measure trace levels of hydrogen for both safe and sustainable energy practices.

Metal-oxide (MOx) sensors have long been recognized for their versatility, cost-effectiveness, and adaptability in gas sensing, with applications in the environmental monitoring of gases such as $CH_4$, $NO_2$, $NH_3$, $CO$, and $CO_2$. Metal oxides can undergo different surface reactions like adsorption-desorption or oxidation-reduction with target gases that lead to measurable changes in the sensor's electrical and/or physical properties.[5–7] Among the extensively studied metal oxides, ZnO stands out as a promising material for practical sensing applications due to its wide band gap, high surface area, chemical stability, affordability, and ease of synthesis.[8–10] However, ZnO faces certain limitations, including response instability, low sensitivity, and poor selectivity, particularly towards hydrogen gas. To address these limitations, several enhancement strategies have been explored, such as noble metal doping (e.g., with Pd, Pt, Au, Sb, Co and Cu) to modify ZnO's electronic properties and improve sensitivity and selectivity by enhancing surface reactivity and tuning charge transfer mechanisms.[10–16] Additionally, surface functionalization has been applied to optimize defect states, such as oxygen vacancies, and to increase adsorption sites, which collectively enhance gas adsorption-desorption kinetics. These approaches have shown promise in improving the overall performance, selectivity, and long-term stability of ZnO-based gas sensors.

Previous studies on ZnO towards hydrogen sensing have often focused on optimising material properties, differentiating the pure and doped material performance, and analysing the effect of crystal size on sensor sensitivity, structural stability, and selectivity.[17–19] For example, Xu et al.[20] quantified the enhanced sensitivity of 0.5 wt% Pt/ZnO nanoparticles, synthesized by chemical precipitation, towards 0.2% hydrogen at an operating temperature of 330 ºC compared to pristine ZnO. Similarly, Rout et al.[15] reported better sensing performance of ZnO nanowires and nanotubes, synthesised by electrochemical deposition and doped with different molar concentrations of Pt, towards hydrogen. Phanichphant et al.[21] produced 0.2-2.0 at. % Pt-doped ZnO by a single-step flame spray pyrolysis and reported successful hydrogen sensing, while Tien et al.[22] deposited Pt-coated ZnO nanorods using molecular beam epitaxy (MBE) and showed enhanced sensitivity towards hydrogen at 500 ppm levels with $N_2$ background at room temperature. Also, Bhati et al.[23] deposited different wt% of Ni-doped (2, 4, and 6% Ni) ZnO nanostructures by radio frequency (RF) magnetron sputtering utilising a Ni chip attached to a ZnO target and found the nanostructures to be highly sensitive towards low concentration levels of hydrogen. Additionally, some studies employed sputtering followed by post-processing heat treatments to achieve hydrogen sensing at lower concentrations.[24,25] Similarly, Hu et al. developed composite Pd-doped ZnO nanostructures with $SnO_2$[17] while Jiao et al. explored the on-chip growth of ZnO with PdO decoration via a chemical synthesis process.[26] Thus, studies so far indicate that ZnO's hydrogen sensing capabilities can be improved via surface decoration of noble metals (such as Pt and Pd), which can be attributed to the noble metals providing catalytic sites that enhance gas adsorption and charge transfer.[21,23] Despite these advancements, the reported limit of detection (LOD) in literature is above 100 ppm and there remains a need for a scalable, reproducible, and cost-effective fabrication technique involving noble metal decorated ZnO that can deliver stable and high-performance sensors.

In this study, we present a simple in-situ sputtering-based technique to fabricate ZnO thin films decorated with Pt, requiring minimal to no additional processing. Our cost-effective and scalable method produces robust and highly sensitive hydrogen sensors, addressing a significant challenge in hydrogen detection research. We also produce durable sensors with high sensitivity and our thin films demonstrate good stability, maintaining consistent responses over one year without any signs of cracks or material degradation. To gain deeper insights into the sensing mechanism, we perform density functional theory (DFT) based calculations, revealing the active role of Pt in the energetics of the hydrogen sensing process on polar ZnO surfaces. By exploring various surface reactions, we identified specific surface terminations and intermediate species that facilitate hydrogen sensing. Thus, our combined computational and experimental approach offers a valuable framework that can be replicated in future gas sensor studies. Notably, our findings highlight the effectiveness of our fabrication process, the fundamental mechanisms that drive sensor performance, and the pivotal role of noble metal decoration in optimizing MOx sensors for practical hydrogen sensing.

## 2. Results

### 2.1 Material Characterisation

The crystal structure of both pristine and Pt decorated ZnO thin films (with Pt deposition times of 1, 2, 4, and 6 sec) was analysed using X-ray diffraction (XRD) in the range of 20° to 80°, as shown in **Figure 1**. All observed peaks align well with the standard JCPDS (Joint Committee on Powder Diffraction Standards) data No. #36-1451, confirming the hexagonal wurtzite structure of ZnO.[27] The diffraction peaks at approximately 34.30°, 47.72°, 55.58°, and 63.13° correspond to the (002), (102), (110), and (103) crystallographic planes, respectively, which are characteristic of wurtzite ZnO.[28] The films demonstrate a strong preferential orientation along the (002) plane, indicating that the ZnO nanostructures grow anisotropically along the *c*-axis. Note that the growth along the *c*-axis is particularly beneficial for gas sensing applications, as it typically enhances surface accessibility and the density of active sites available for gas adsorption.[29] The increased peak intensity along the (002) plane further suggests a high crystalline quality, which is advantageous for consistent electron transport and response stability in gas sensing. No Pt-related diffraction peaks are detected in the Pt decorated ZnO films, likely due to the minimal Pt content, which does not form distinct crystalline phases that are detectable by XRD. The crystallite size of the ZnO films was calculated from the XRD data and is discussed in the supporting information (SI, see **Table S1**).

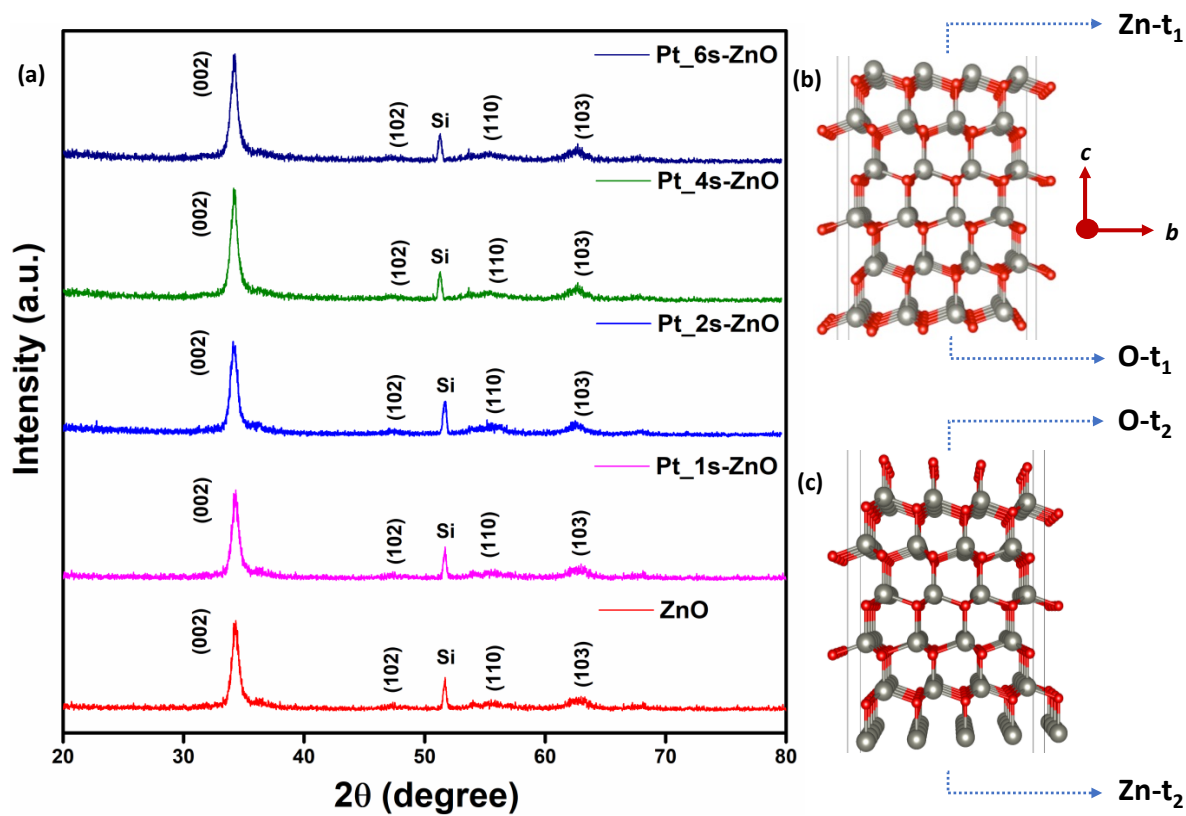

**Figure 1.** (a) XRD spectra of pure (red line) and Pt decorated ZnO thin films on Si substrates. Pink, blue, green, and violet lines indicate Pt sputtered for 1, 2, 4, and 6 sec on top of the ZnO, respectively. (b and c) Two possible terminations of the ZnO (002) surface, namely $t_1$ (panel b) and $t_2$ (panel c). Red atoms are oxygen, grey atoms are zinc, and solid black lines are cell boundaries. Both terminations have sides that are filled with exposed Zn (labelled Zn-$t_1$ and Zn-$t_2$) and exposed O (O-$t_1$ and O-$t_2$) atoms.

The (002) surface of ZnO, which we observe to be the preferred direction of growth from our XRD measurements (**Figure 1**a), is a polar surface that can have two possible terminations each with exposed Zn and O atoms, as shown in **Figure 1**b and c, i.e., the $t_1$ termination with exposed Zn and O (Zn-$t_1$ and O-$t_1$, **Figure 1**b) and the $t_2$ termination with exposed Zn and O (Zn-$t_2$ and O-$t_2$, **Figure 1**c). Multiple studies on ZnO-based sensors have indicated the polar planes (001) and (002) to be the dominant plane(s) upon synthesis, and have been considered to contribute heavily in the sensing performance via the presence of active sites.[30–34] We employed DFT to calculate the absolute surface energies of both terminations of the (002) plane to determine the ground state configuration (see **Table S6** and **Section S5** of the SI for details). Since both terminations of the (002) surface is polar with a net dipole moment perpendicular to the surface along the $c$-axis, we capped one side of the surface with pseudo-hydrogen atoms (see **Figure S4** and **Table S5**)[35–37] and used the non-polar (100) ZnO surface as the reference surface to compute absolute surface energies. Notably, the O-$t_1$ (Zn-$t_1$) and O-$t_2$ (Zn-$t_2$) surfaces exhibit surface energies of 1.84 (1.90) and 9.57 (5.83) J/m$^2$, respectively, indicating that the $t_1$ termination is thermodynamically favored to form. Thus, we expect one of the $t_1$ terminations to form primarily during sputtering and use both Zn and O exposed facets of the $t_1$ termination to perform further DFT calculations.

We employed scanning electron microscopy (SEM) to investigate the surface morphology and structural integrity of pristine and Pt decorated ZnO thin films. Panels a-e in **Figure 2** illustrate that the films exhibit a smooth, wrinkle-free surface without any visible cracks, even after Pt decoration. Such uniform distribution across the substrate suggests high-quality thin-film deposition with minimal defects, which is essential for reliable sensor performance. Cross-sectional SEM imaging reveals a progressive increase in film thickness corresponding to increased Pt deposition time. For example, the thickness for pristine ZnO is measured at 39.8 nm (**Figure 2**a), while the thickness for Pt decorated ZnO is around 40 nm, 40.2 nm, 41.4 nm, and 43.2 nm for 1, 2, 4, and 6 sec of Pt deposition, respectively (**Figure 2**b–e). Thus, our in-situ process with controlled Pt deposition allows for precise tuning of the film's sensing properties, unlike chemical techniques. **Figure 2**f shows SEM imaging of the interdigitated electrodes (IDE), where Pt decorated ZnO particles are evenly distributed, confirming the homogeneity of the sputtering deposition process. This homogeneity is crucial in sensing devices, as it ensures that the entire sensing region of the IDE is covered with active material, allowing for consistent response when exposed to target analytes.

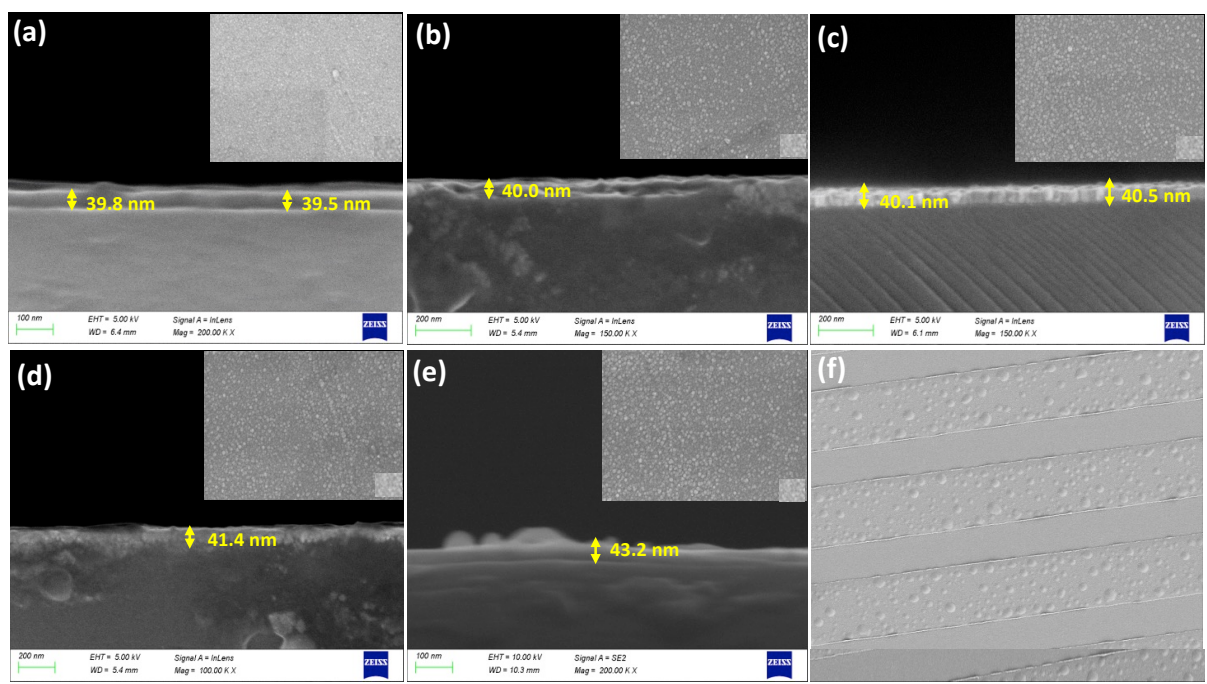

**Figure 2.** Field emission SEM (FESEM) morphology and thickness from cross-sectional imaging of (a) pristine ZnO and Pt decorated ZnO decorated with (b) 1 sec (c) 2 sec (d) 4 sec and (e) 6 sec deposition times. (f) SEM image of uniform coverage of Pt decorated ZnO thin film (at 2 sec deposition time) over the IDE.

Electron probe microanalysis (EPMA) was conducted to examine the elemental distribution within pristine and Pt decorated ZnO thin films. **Figure S1** provides an elemental map of Pt decorated ZnO (at a deposition time of 2 sec), highlighting the distribution of Si, Zn, O, and Pt across the IDEs. We performed EPMA point analysis was for Pt deposition times of 1, 2, 4, and 6 sec, and summarized the results in **Table S2**. Importantly, we observe a progressive increase in Pt content as deposition time is extended, ranging from 1.30 to 14.62 mass% Pt from 1 sec to 6 sec deposition time, indicating gradual enhancement of Pt coverage on the ZnO surface.

To get a deeper understanding of the chemical oxidation states and elemental composition of the thin films, we performed X-ray photoelectron spectroscopy (XPS) on all the fabricated ZnO films. As shown in **Figure 3**a, Zn and O are present in the pristine ZnO film, and an additional peak corresponding to Pt is observed in the Pt decorated ZnO films (highlighted in yellow). No additional peaks for other impurities were detected, confirming the high purity of the fabricated samples. For both pristine and Pt decorated ZnO films, the XPS spectra exhibit two prominent peaks at binding energies of 1045.2 eV and 1022.2 eV, which correspond to the Zn 2p1/2 and Zn 2p3/2 levels, respectively. These peaks are characteristic of the +2-oxidation state of zinc ($Zn^{2+}$) that is expected within the ZnO structure, as shown in **Figure 3**b. The core-level XPS spectra of Pt 4f in Pt decorated ZnO films, presented in **Figure 3**c, highlighting two distinct doublet peaks across all samples, which we identify as Pt 4f7/2 and Pt 4f5/2 at binding energies of 70.8 eV and 74.8 eV, respectively. Both peaks correspond to the $Pt^0$ oxidation state, which indicates metallic Pt.[38] Additionally, the O 1s core-level spectra in Pt decorated ZnO films, shown in **Figure 3**d, exhibit two primary components: lattice oxygen species ($O_{lat}$) at ~530.5 eV and adsorbed oxygen species ($O_{ads}$) at ~531.5 eV. The relative amounts of $O_{lat}$ and $O_{ads}$ are calculated and presented in **Table S3**. Notably, the Pt decorated ZnO films have a higher relative percentage of $O_{ads}$ compared to pristine ZnO, where the $O_{ads}$ can contribute better to gas sensing than $O_{lat}$.[39,40] In summary, the XPS data indicates the presence of both metallic Pt and higher amount of $O_{ads}$ that can act in a combined manner and give rise to better hydrogen sensing response.

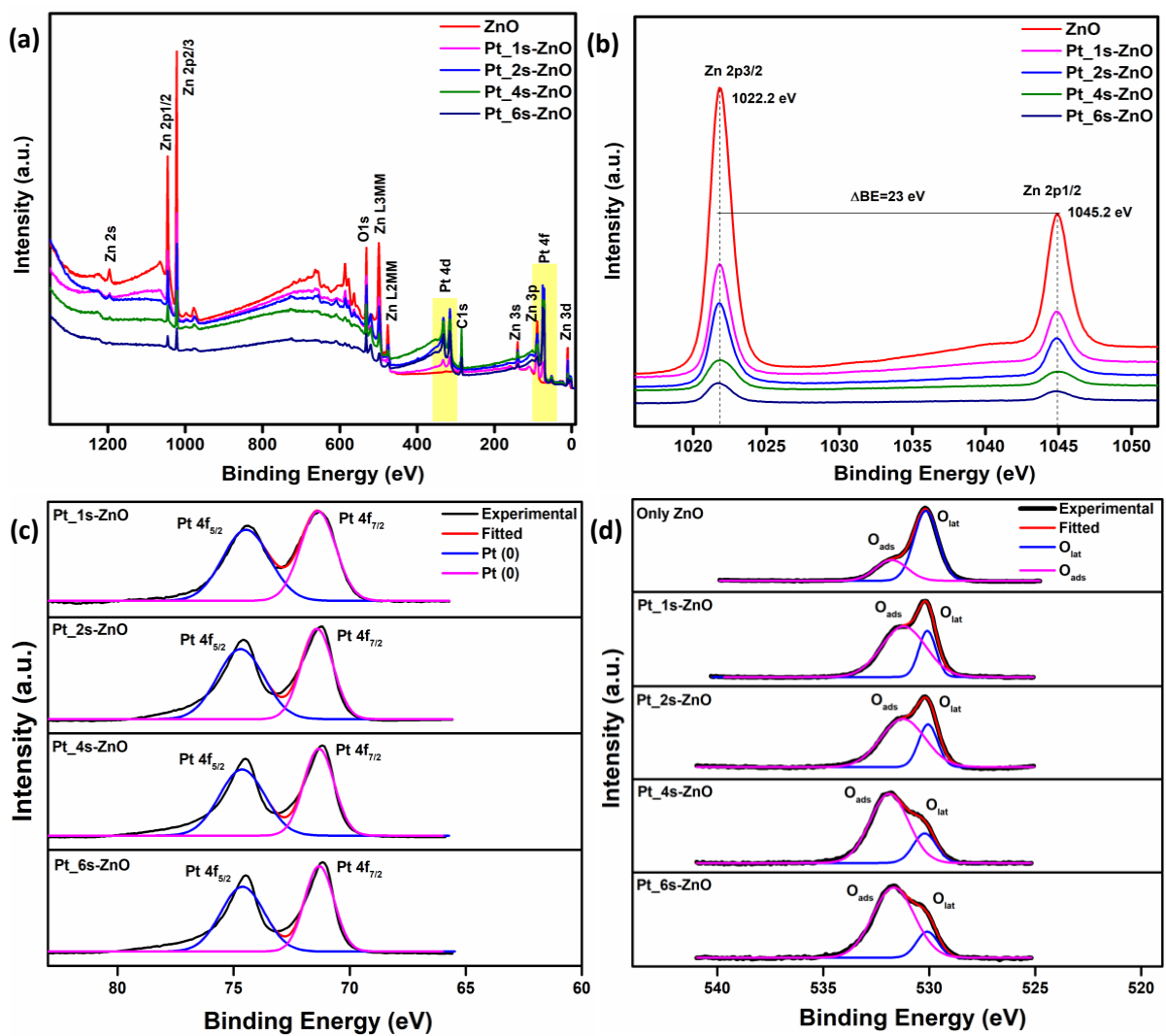

**Figure 3.** (a) Overall XPS spectra and (b) Zn 2p section of the XPS spectra in pristine (red) and Pt decorated ZnO thin films, where pink, blue, green, and violet lines indicate Pt deposition times of 1, 2, 4, and 6 sec, respectively. (c) Pt 4f and (d) O 1s core-level XPS spectra of ZnO thin films. While the Pt 4f spectra are displayed only for Pt decorated samples, the O 1s spectra are shown for both pristine and Pt decorated ZnO films. The panel labels indicate Pt decoration times (1, 2, 4, and 6 sec).

For analysing the surface roughness and topological characteristics of the ZnO thin films, we used atomic force microscopy (AFM). Particularly, we investigated the change in root mean square (RMS) roughness to understand the effect of surface roughness on gas sensing performance. AFM images in both two-dimensional (2D) and three-dimensional (3D) views across a 1 μm scanning area, as displayed in panels a-f of **Figure 4**, demonstrate a uniform film coverage across the substrate. Notably, quantitative analysis reveals that surface roughness increases with increasing Pt deposition time, with RMS values measured at ~0.79 nm for pristine ZnO, and subsequently increasing to 1.23 nm, 1.98 nm, 2.15 nm, and 2.69 nm for ZnO films decorated with Pt for 1, 2, 4, and 6 sec, respectively. The progressive roughening of the surface due to Pt decoration is visible in the AFM images (see panels b and e of **Figure 4**), where the ZnO surface layer becomes increasingly textured compared to the smooth surface of pristine ZnO, possibly allowing for more sites for hydrogen chemisorption in the Pt decorated sample.[41,42] The higher roughness can also facilitate the adsorption of oxygen species, which may be crucial for hydrogen sensing, given the possible reactivity of adsorbed oxygen towards adsorbed hydrogen. The AFM data also shows a clear phase difference between the pristine and Pt decorated ZnO thin films (panels c and f of **Figure 4**), further highlighting the topological changes caused by the Pt decoration.

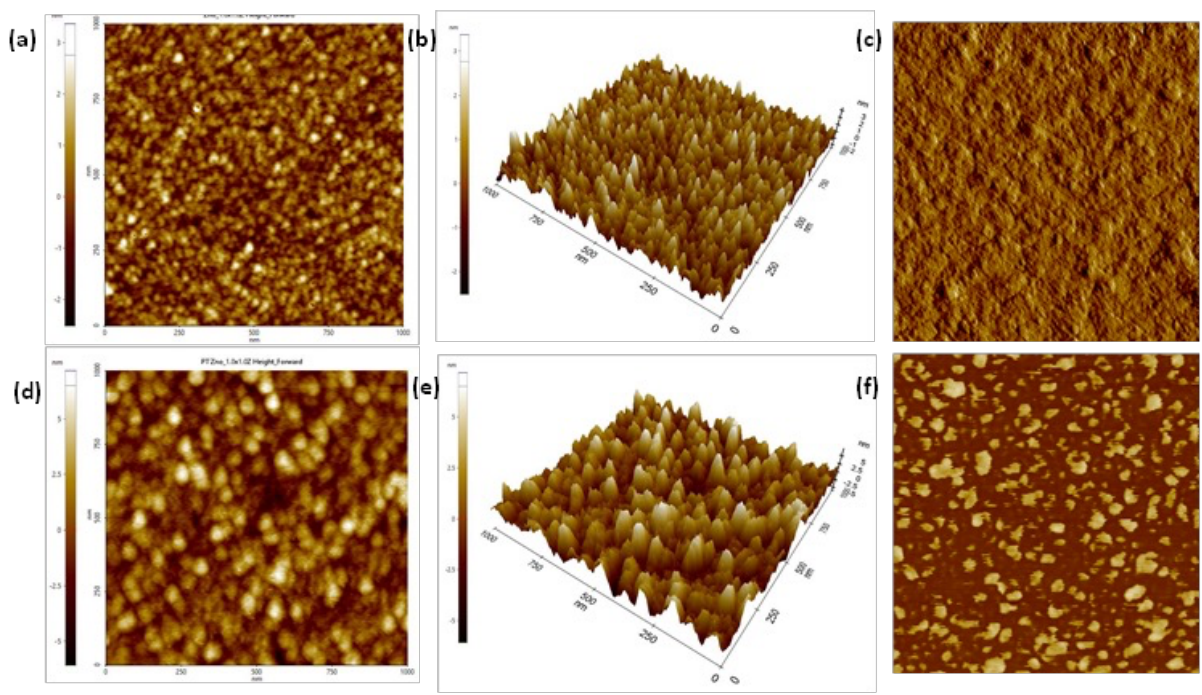

**Figure 4.** (a, d) 2D, (b, e) 3D, and (c, f) phase difference images obtained from AFM of pristine (top row) and Pt decorated (bottom row) ZnO thin films. We performed AFM measurements on the Pt decorated sample with 2 sec of Pt deposition.

We conducted temperature-dependent current-voltage (I-V) measurements across a -1 V to +1 V range to evaluate the suitability of the IDEs for hydrogen sensing (see **Figure S2**). The devices based on pristine ZnO and Pt decorated ZnO with deposition times of 1, 2, and 4 sec display typical semiconducting behaviour between 25 °C and 200 °C, with an increase in current as a function of increasing temperature, which is desirable for gas sensing. In contrast, the ZnO device with 6 sec of Pt deposition exhibits a decrease in current with increasing temperature, signifying metallic behaviour. Thus, the excessive Pt content in the 6 sec sample creates a conductive path for electrons that overrides the underlying semiconducting response of ZnO, reducing the sensitivity of the sample to any interactions with gas molecules, and making the sample unsuitable for sensing (**Figure S2**). Therefore, we select pristine and Pt decorated ZnO (with 1, 2, and 4 sec of Pt deposition) for further hydrogen sensing experiments.

## 2.2 Gas sensing response

To evaluate the gas sensing performance, we exposed all fabricated sensors to hydrogen gas (of different concentrations) at varying operating temperatures, ranging from 25 °C to 300 °C. The sensing response of the material towards different hydrogen concentrations is quantified as a response (in %) using **Equation 1**, where $I_g$ and $I_a$ correspond to the measured current in the presence of hydrogen gas and synthetic air, respectively.

$$\text{Response}(\%) = \frac{I_g - I_a}{I_a} \times 100\% \quad \ldots (1)$$

**Figure 5**a plots the response of pristine ZnO (red lines and symbols) and Pt decorated ZnO with 1 sec (pink), 2 sec (blue), and 4 sec (green) Pt deposition times as function of operating temperature. Notably, pristine ZnO exhibits its highest response at 275 °C, while the Pt decorated sensors achieve their maximum response at 225 °C. Typically, sensors exhibit an optimal operating temperature that is determined by the trade-off between the spontaneity of gas species adsorption on the surface and any activation energies that the system has to overcome to facilitate a chemical reaction (or a change in material property). Importantly, Pt decorated samples (with 1 sec and 2 sec deposition times) display significantly higher response (~11.5 and 23.5 times higher than pristine ZnO), highlighting the effectiveness of the Pt decoration over ZnO in hydrogen sensing. Given that Pt decorated samples exhibit their highest response at ~225 °C, we select this as the optimal temperature for further sensing measurements.

**Figure 5**b displays the response (plotted as measured current) of pristine and Pt decorated ZnO based sensors as a function of exposure time to 10k parts per million (ppm) of hydrogen gas operated at their optimal operating temperature with an exposure time of 3 min. Importantly, on the Pt decorated ZnO sensor (with 2 sec deposition time) the response to hydrogen gas reaches 52,987% which is significantly higher over other samples including the 1 sec Pt deposited ZnO (25,870%), 4 sec Pt-deposited ZnO (475%), and pristine ZnO (2,250%). All the sensors demonstrate reversibility upon switching the hydrogen gas to the baseline dry air. Thus, the presence of Pt on ZnO modifies the electronic properties of ZnO to amplify the sensor's response by facilitating electron exchange with hydrogen. However, the reduction in response at higher degrees of Pt coverage (as is obtained with the 4 sec Pt-deposited ZnO sensor) is possibly due to an increased metallic character of the surface that suppresses the underlying semiconducting nature of ZnO and possible excessive coverage of the active sites on ZnO by Pt.

Given our response measurements (panels a and b of **Figure 5**), we find the optimal Pt deposition time to be 2 sec on the pristine ZnO surface that is able to balance the increased interactions with hydrogen gas without adversely affecting the semiconducting nature of ZnO. Indeed, we find the 2 sec Pt deposited sample to show the best response among other samples at operating at 225 °C across a range of hydrogen concentrations (i.e., from 100 to 10k ppm), as shown in **Figure 5**c. Additionally, we find the 2 sec Pt deposited sample to exhibit excellent stability in our dynamic response measurements (**Figure 5**d) while operating at 225 °C. Specifically, we find the Pt decorated sensor to show monotonically decreasing response as hydrogen concentration is varied from 10k to 1k ppm. Moreover, the samples show high degree of reversibility by recovering the baseline response whenever the hydrogen supply is shut off during the measurement and display highly reproducible behaviour across multiple batches of sensors that were fabricated (see **Figure 8**c). Thus, we find the performance of our 2 sec Pt-deposited ZnO sensor to be highly promising for hydrogen gas detection at ppm levels.

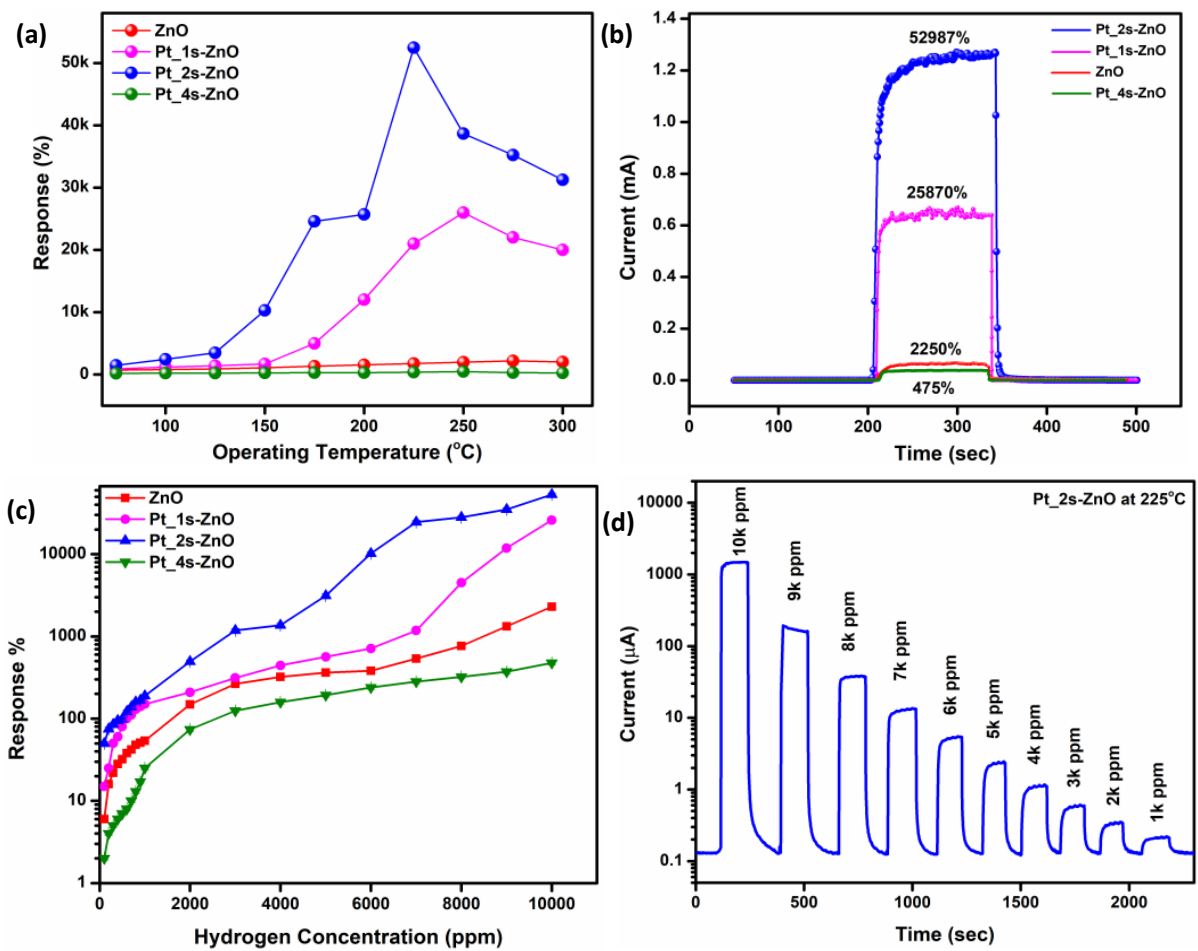

**Figure 5.** (a) Response (%) variation with operating temperature varying from 25 °C to 300 °C at 10k ppm of hydrogen gas, (b) response (measured in current) as a function of time when 10k ppm of hydrogen gas is introduced, and (c) response at different hydrogen concentration levels for pristine and Pt decorated ZnO thin film sensors. Red, pink, blue, and green symbols and lines in panels a, b, and c represent pristine, 1 sec, 2 sec, and 4 sec Pt deposited ZnO samples. (d) Dynamic response of the 2 sec Pt-deposited ZnO sample at 225 °C towards hydrogen concentrations varying from 10k ppm to 1k ppm over time.

To understand the differences in sensing response between the pristine and Pt decorated ZnO, we investigated $H_2$ adsorption on both surfaces using DFT. We primarily used binding energies ($E_{binding}$), as defined in **Equation 2**, to examine the spontaneity of adsorption of any species over the ZnO surface. $E_{slab+adsorbate}$, $E_{slab}$, and $E_{adsorbate}$ indicate the DFT-calculated total energies of the slab model along with an adsorbate, the pristine slab model, and the isolated adsorbate, respectively. The adsorbate here can be a molecule and/or a cluster of Pt atoms on the ZnO surface. In general, negative binding energies indicate strong spontaneous adsorption of a species on a slab (or substrate).

$$E_{binding} = E_{slab+molecule} - E_{slab} - E_{molecule} \qquad \ldots (2)$$

Given that Pt decoration plays a crucial role in sensor performance, we placed a 4 atom Pt cluster (or $Pt_4$ cluster) on the pristine ZnO surface to capture the effect of Pt. Although larger Pt clusters generally more stable,[43,44] we considered a 'small' cluster in our work to capture the effect of Pt at a reduced computational cost. We calculated binding energies for the planar and tetrahedral geometries of the $Pt_4$ cluster (i.e., adsorbate in **Equation 2** is a $Pt_4$ cluster), initialised on different sites on both the O-$t_1$ and Zn-$t_1$ surfaces, with the low energy relaxed geometries displayed in **Figure S5** and the lowest energies compiled in **Table S7**. Importantly, we find the planar $Pt_4$ geometry to be the stable configuration on both O-$t_1$ and Zn-$t_1$ surfaces and subsequently use this configuration for further $H_2$ adsorption calculations (see below).

For modelling $H_2$ adsorption on the pristine and Pt decorated ZnO surface, we considered four distinct surface sites: Zn, O, hollow, and Pt, as shown in **Figure 6**a. Note that the on-Pt site is the 'point-of-difference' between the pristine and Pt decorated ZnO surfaces, i.e., we consider the on-Pt site to be the only additional site that is available upon Pt decoration while the three remaining sites are already available for $H_2$ adsorption in pristine-ZnO. Thus, any changes that Pt decoration can effect on ZnO, with respect to $H_2$ adsorption, will be captured by the binding energy exhibited by the on-Pt site compared to the three remaining sites.

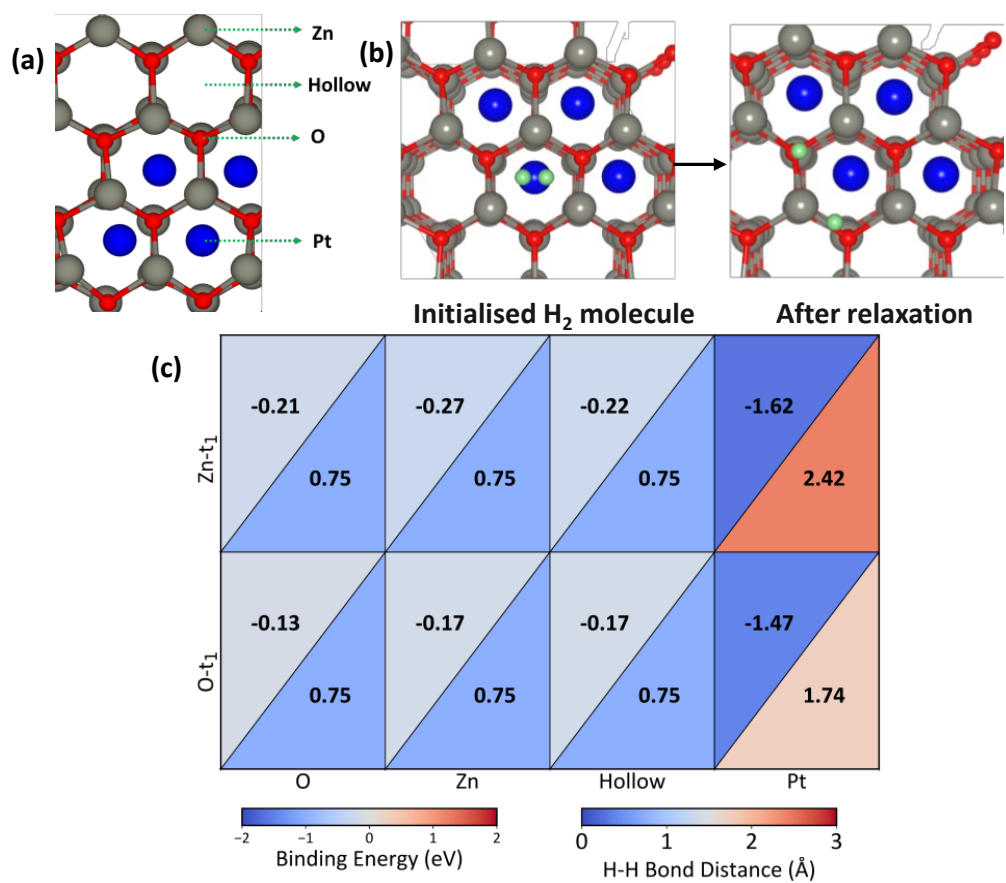

**Figure 6.** (a) Top view of four adsorption sites considered for possible $H_2$ adsorption are illustrated on a Zn-$t_1$ ZnO surface that is decorated with a $Pt_4$ cluster. (b) Demonstration of facile dissociation of $H_2$ molecule on the Pt cluster upon structure relaxation. Blue, grey, red, and green spheres as Pt, Zn, O, and H, respectively. (c) $H_2$ binding energies (in eV, upper triangles) and H-H bond distances (in Å, lower triangles) post-adsorption for various initialisations (on O, on Zn, hollow, and on Pt sites) on the Zn-$t_1$ (upper row) and O-$t_1$ (lower row) surfaces of ZnO.

The binding energies for $H_2$ adsorption on each of the four sites on the O-$t_1$ and Zn-$t_1$ surfaces are presented as a heat map in **Figure 6**c, where the upper triangle within each cell represents the corresponding binding energy. While considering $H_2$ adsorption on a Pt decorated cluster, we refer to $H_2$ molecule as the adsorbate in **Equation 2** and the ZnO surface including the $Pt_4$ cluster as the slab. The lower triangles in each cell of **Figure 6** indicate the H-H bond distance post-adsorption (i.e., upon structure relaxation). We observe that $H_2$ binds quite weakly, i.e., without significant dissociation of the $H_2$ molecule, on pristine ZnO surfaces. For example, the $H_2$ binding energies on both O-$t_1$ and Zn-$t_1$ range from –0.13 to –0.22 eV across the on O, on Zn, and hollow sites (**Figure 6**c), with the $H_2$ bond distances post-adsorption (~0.75 Å) quite similar to the equilibrium bond length of an isolated $H_2$ molecule (0.74 Å). In contrast, $H_2$ adsorbs strongly on Pt decorated ZnO, as indicated by binding energies on Pt site of –1.47 eV and –1.62 eV on the O-$t_1$ and Zn-$t_1$ surfaces, respectively, along with spontaneous dissociation post-adsorption (H-H bond distances of 1.74 to 2.42 Å, see **Figure 6**b and c). Upon dissociation, the atomic H form bonds with the Pt atoms at a bond lengths ranging from 1.6-1.7 Å, with **Figure 6**b illustrating the spontaneous dissociation of $H_2$ on the Pt cluster during a DFT structure relaxation calculation. Thus, we expect Pt decoration to significantly facilitate both $H_2$ adsorption and $H_2$ dissociation on the ZnO surface, thereby enhancing the reactivity of hydrogen with available oxygen and reducing any kinetic barriers associated with $H_2$ dissociation.

## 2.3 LOD, Selectivity, response, recovery, and stability

To evaluate the lower detection limit of the 2 sec Pt-deposited sensor (at 225 °C), we measured the response over a range of low hydrogen concentrations, from 9.7 ppm to 0.1 ppm. The dynamical response of the sensor, plotted as measured current, is displayed in **Figure 7**a and demonstrates a strong response of ~1689% at 9.7 ppm hydrogen. Importantly, the sensor yields a measurable 38% response at the extremely low hydrogen concentration of 0.1 ppm, which we identify as the limit of detection (LOD), highlighting the ability of the sensor to detect hydrogen leaks early and fast even under low (or trace) concentrations. Moreover, the sensor exhibits a linearly varying response (dotted red line shown in **Figure S3**a) at low hydrogen concentrations, which simplifies calibration of the device in a practical application. Also, we observe marginal hysteresis in the dynamical response as hydrogen gas concentration is increased and decreased (**Figure 7**a), which may be attributed to differences in adsorption and desorption kinetics. Similarly low levels of hysteresis are observed for the Pt decorated sample even across larger ranges of hydrogen concentrations (100 to 10k ppm and vice-versa, see **Figure S3**b). Thus, the Pt decorated ZnO thin film is a promising framework for hydrogen sensing as it combines a swift and reversible response with an extremely low LOD.

The response and recovery characteristics of the sensor, as depicted in panels b and c of **Figure 7**, provide critical insights into its dynamic performance when exposed to hydrogen. The response time ($t_{res}$) is defined as the duration needed for the sensor to achieve 90% of its maximum current after exposure to hydrogen gas, which reflects the sensor's ability to quickly detect changes in the target gas concentration. Conversely, the recovery time ($t_{rec}$) is the time the sensor takes to return to 90% of its baseline current after hydrogen is removed and synthetic air is introduced. As shown in **Figure 7**, the Pt decorated ZnO (with 2 sec deposition time) sensor exhibits significantly faster response and recovery times ($t_{res}$=10 sec, $t_{rec}$=3 sec, **Figure 7**c) compared to pristine ZnO ($t_{res}$=45 sec, $t_{rec}$=22 sec, **Figure 7**b) for 10k ppm of hydrogen concentration, highlighting the enhanced kinetics of adsorption and desorption provided by Pt decoration. Thus, we confirm that Pt decoration of ZnO significantly improves $t_{res}$ and $t_{rec}$ compared to pristine ZnO, showcasing the critical role of Pt in enhancing the sensor's performance.

Selectivity is a crucial parameter for gas sensors, as it defines their ability to specifically detect a target gas in the presence of other ambient gases. To evaluate selectivity, we tested pristine and Pt decorated (2 sec Pt-deposition) ZnO sensors with various gases under identical measurement conditions, including operating temperature. Notably, both pristine and Pt decorated ZnO sensors demonstrated higher selectivity for hydrogen over other gases such as CO (7.16 ppm), ammonia (5.2 ppm), ethylene (5.2 ppm), methane (10 ppm), $NO_2$ (5 ppm), acetone (10.5 ppm), $CO_2$ (1026 ppm), and $O_2$ (10k ppm), as quantified by the higher response shown towards hydrogen in **Figure 7**d. Given that the incorporation of Pt onto ZnO enhances the catalytic activity of the sensor surface by favoring the adsorption and dissociation of hydrogen molecules (**Figure 6**), we do observe enhanced selectivity of the Pt decorated ZnO sensor towards hydrogen compared to pristine ZnO, as quantified by a ~4-fold increase in the response (**Figure 7**d).

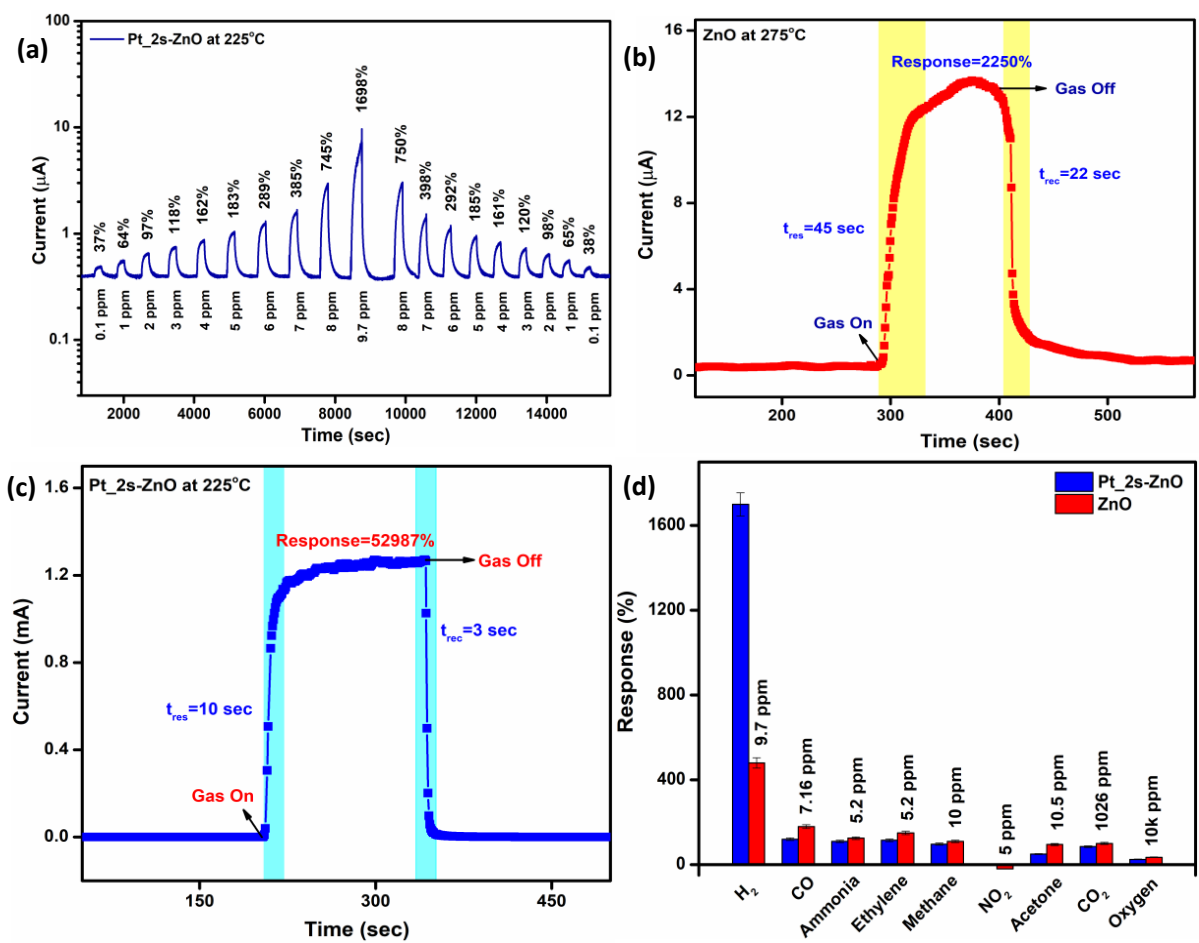

**Figure 7.** (a) Dynamical response of Pt decorated ZnO (2 sec deposition time, 225 °C operating temperature) at low hydrogen concentrations ranging from 0.1 to 9.7 ppm. (b and c) Response and recovery time quantification at 10k ppm hydrogen concentration, as highlighted by the yellow and blue highlighted regions, for (b) pristine and (c) Pt decorated (2 sec deposition time) ZnO sensor, operating at 275 °C and 225 °C, respectively. (d) Selectivity of pristine (red bars) and Pt decorated (blue, 2 sec deposition time) ZnO sensors towards different gases, as quantified by the corresponding response (in %).

The repeatability of the sensor response in detecting hydrogen at a 1% concentration was assessed across multiple cycles for both pristine and Pt decorated (2 sec deposition time) ZnO sensors. As quantified by the measured current in panels a and b of **Figure 8**, both pristine and Pt decorated ZnO sensors exhibit consistent response across cycles. Specifically, the Pt decorated sensor shows particularly consistent response (~51,825% to 52,987%, **Figure 8**b) compared to the pristine sample (~1,979% to 2,250%, **Figure 8**a). This consistency over multiple cycles indicates reliable sensor functionality with repeated hydrogen exposures, particularly for the Pt decorated sensor. Additionally, the reproducibility of the sensors was verified by comparing the response across three independently fabricated samples in different batches. As shown in **Figure 8**c, the response variation across these sensors was minimal, indicating high reproducibility when exposed to different hydrogen concentrations. To evaluate long-term stability, we tracked the sensor response over several days, for both the pristine (red symbols) and Pt decorated (blue symbols) samples, as shown in **Figure 8**d. Notably, the Pt decorated ZnO sensor maintains a stable response after approximately one year of fabrication, which is likely due to the robust chemical and structural integrity of the Pt-ZnO interface. On the other hand, the response of pristine ZnO drops steadily with time after fabrication. Thus, the durability and long-term stability of the Pt decorated ZnO sensor's performance further underscores its potential for practical hydrogen detection.

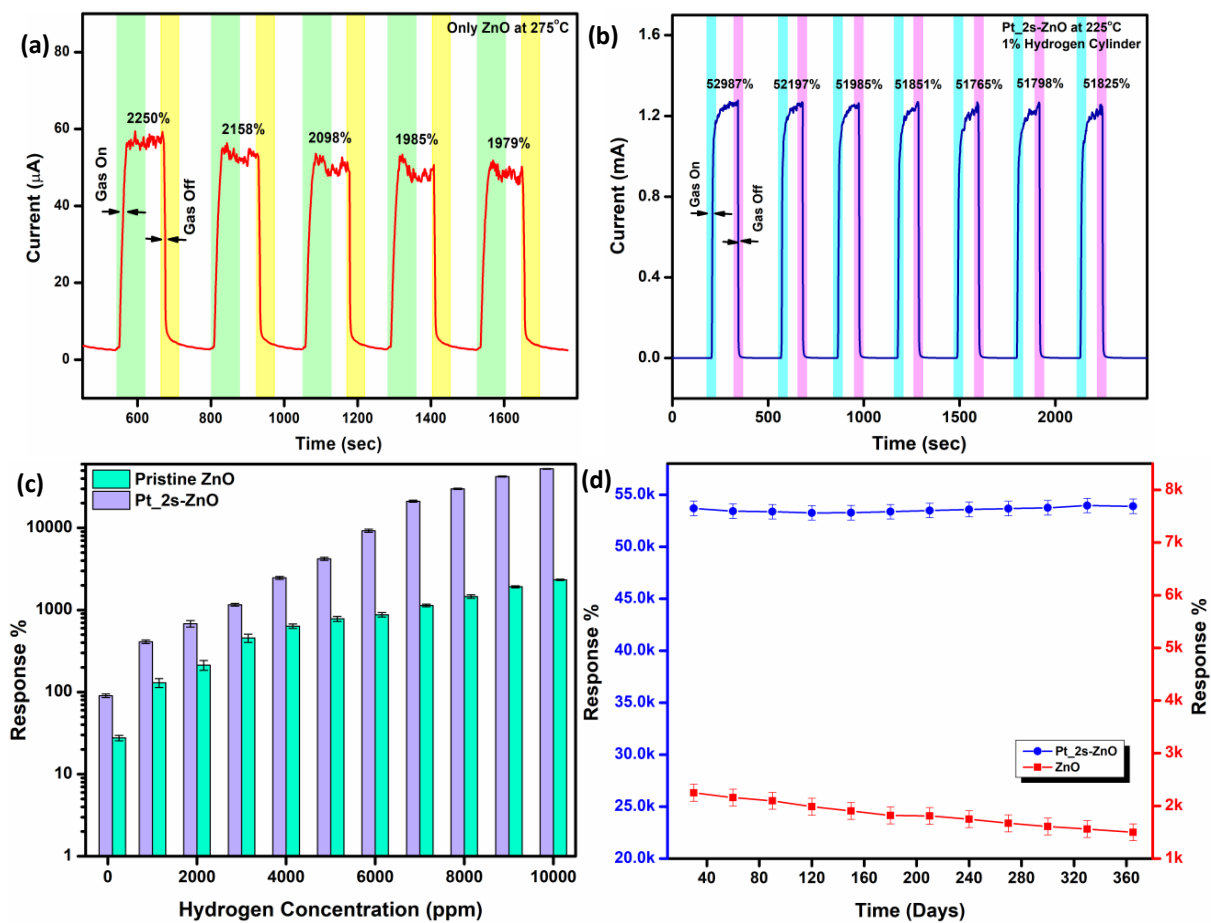

**Figure 8.** Repeatability in response over multiple cycles of 1% hydrogen exposure for (a) pristine and (b) Pt decorated (2 sec deposition time) ZnO, operating at 275 °C and 225 °C, respectively. (c) The average response % over three samples as a function of loaded hydrogen concentration, with the error bars representing the standard deviations. (d) Long-term stability in sensor response of the pristine ZnO (red symbols) and Pt decorated ZnO (2 sec deposition time, blue symbols) devices with time after fabrication.

In gas sensing applications, the work function is a key parameter that helps to understand the fundamental mechanisms of charge transfer and reaction kinetics. It directly influences how a material interacts with gas molecules, as the adsorption and desorption of gases involve electron transfer processes that affect the material's surface potential.[45,46] To investigate these mechanisms, we conducted Kelvin probe force microscopy (KPFM) measurements on both ZnO and Pt films to calculate their work function values. In this study, we derived the work function values of ZnO and Pt films using the **Equation S3** (see **Section S3** of SI). We found the work function of ZnO to be approximately 4.5 eV, while that of Pt was 5.2 eV. Due to this difference in the work function values, electrons from the surface of ZnO will transfer to the Pt atoms until the Fermi levels of both materials equilibrate. The electron transfer can lead to the formation of a depletion region at the ZnO-Pt junction, where the semiconductor band edges bend 'upwards' with consequent impact on the sensor's electrical conductivity and response to gas exposure.

**2.4 Gas sensing mechanism**

We performed DFT-based calculations quantifying adsorption and reaction energetics on pristine and Pt decorated ZnO surfaces to unearth the gas sensing mechanism and the specific ZnO surface termination that is contributing to our measured response, to further understand the intrinsic response of ZnO and the role of Pt in hydrogen detection. Within the gas sensing literature of metal oxides, the widely accepted mechanism is the 'adsorbed oxygen' model,[47,48] where the oxygen from the atmosphere gets adsorbed on the surface of the oxide sensor under ambient conditions. The adsorbed oxygen pulls out electrons from the metal oxide thereby reducing the free electron concentration for an *n*-type semiconductor (or increasing the free hole concentration for a *p*-type semiconductor). When the oxide surface is subsequently exposed to $H_2$, the adsorbed oxygen reacts to form water, thereby releasing the electrons that were captured by the oxygen and causing an increase in measured current for *n*-type semiconductors (decrease in current for *p*-type semiconductors). This change in measured current is the sensor's response to the gas' (hydrogen's) presence. Similar to forming $H_2O$, the adsorbed oxygen can also react with $H_2$ to form OH, in which case the number of electrons released back is fewer compared to $H_2O$ formation causing a reduced change in measured current. However, there has been recent evidence of tangible sensor response to a target gas (such as $H_2$) in the absence of ambient oxygen along with evidence of poor response in the presence of ambient oxygen,[49] thereby contesting the adsorbed oxygen model. In such cases, the sensor response can be attributed to the Mars van Krevalan mechanism,[50,51] where the target ($H_2$) gas plucks out one of the oxygen from the metal oxide lattice, leaving behind electrons that causes a change in measured current. The lattice oxygens removed by the target gas can subsequently be replenished by atmospheric oxygen under ambient conditions (and/or at elevated temperatures), after the target gas has been flushed out. Thus, it is important to identify the specific mechanism that is in play in our ZnO based device which will facilitate further improvements and optimizations of performance.

The binding energies of various adsorbates, including atomic and molecular hydrogen, atomic and molecular oxygen, and $H_2O$ on both the O-$t_1$ (panel a) and Zn-$t_1$ (panel b) surfaces of ZnO are compiled in **Figure 9**. We considered four possible sites for adsorption of all species, identical to **Figure 6a**, wherein the Pt site represents the adsorption/binding of species in the presence of the Pt cluster on the ZnO surface. On the pristine O-$t_1$ surface, we find atomic hydrogen to bind strongly (~-2.37 to –2.34 eV) compared to molecular $H_2$ (~-0.17 to –0.13 eV) on all sites. Atomic hydrogen adsorption is also stronger on the O-$t_1$ than the Zn-$t_1$ surface (-2.37 vs -0.72 eV). On the other hand, both atomic (-0.99 and –0.84 eV) and molecular (-1.62 and –1.47 eV) hydrogen bind strongly on the Pt site on both Zn-$t_1$ and O-$t_1$ surfaces, highlighting Pt's role as a facilitator of $H_2$ adsorption on the ZnO surface. Moreover, we observe spontaneous dissociation of molecular $H_2$ on the Pt site (see **Figure 6b**) to form atomic hydrogen, which can favour the reaction of hydrogen towards oxygen. Thus, the facilitation of molecular hydrogen adsorption and its dissociation to the reactive hydrogen form is one of the contributing factors to our observation of better sensor response on the Pt decorated ZnO devices compared to the pristine ZnO sensors.

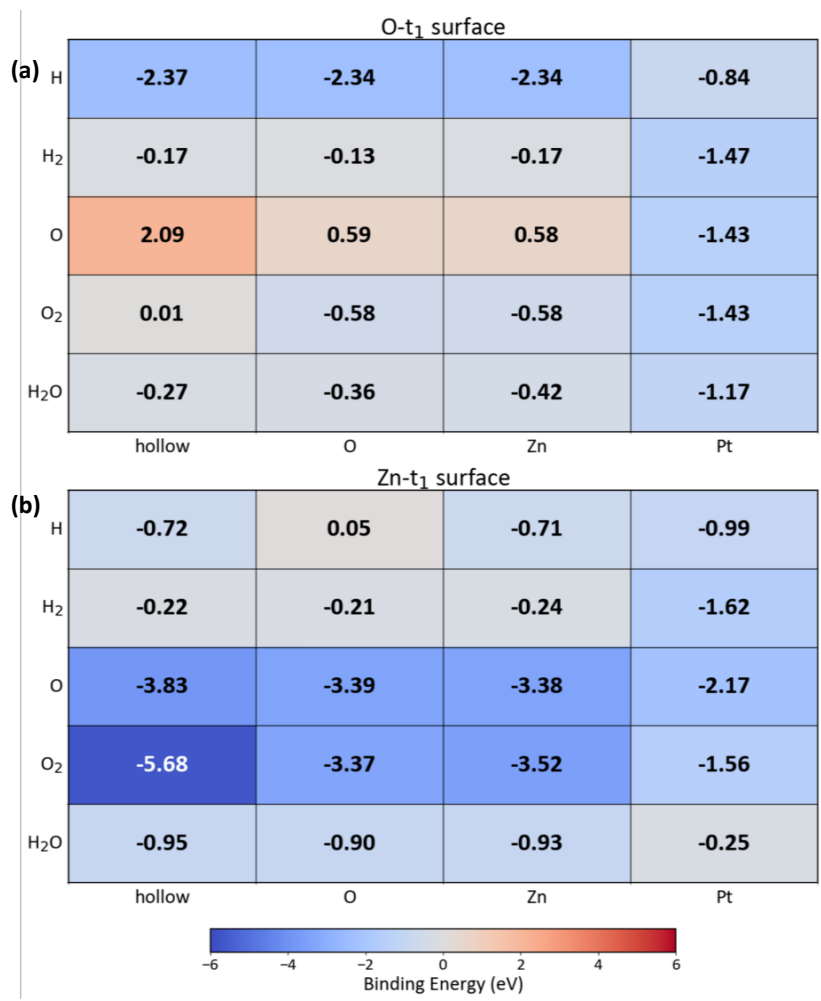

**Figure 9.** Heat map of binding energies of atomic H, molecular $H_2$, atomic O, molecular $O_2$, and molecular $H_2O$ on the (a) O-$t_1$ and (b) Zn-$t_1$ surfaces. The hollow, O, Zn, and Pt correspond to possible sites for adsorption species, as defined in **Figure 6**.

While molecular $O_2$ can adsorb on the O and Zn sites on O-$t_1$ (-0.58 eV, **Figure 9**), atomic oxygen does not bind on the O-$t_1$ site (~0.58 to 2.09 eV), highlighting the non-spontaneity of oxygen dissociation on the O-$t_1$ surface. Thus, we don't expect the adsorbed oxygen pathway to be active on the O-$t_1$ surface in pristine ZnO. On the other hand, Pt decoration enables the adsorption of both molecular and atomic oxygen (-1.43 eV), indicating that Pt decoration can activate the adsorbed oxygen pathway on the O-$t_1$ surface, thereby possibly improving sensing response. In the case of the Zn-$t_1$ surface, oxygen adsorption, both in molecular and atomic forms, is highly favoured (binding energies below –3.3 eV), indicating that the adsorbed oxygen mechanism can be active on the pristine Zn-$t_1$ surface. Particularly, the $O_2$ molecule binds strongly at the hollow site on Zn-$t_1$ (-5.68 eV). The molecule dissociates, and each O atom occupies a hollow position on the surface to form 3 distinct bonds with nearby Zn atoms during a DFT structure relaxation. Pt decoration on Zn-$t_1$ does increase the binding energies of both atomic and molecular oxygen (by a minimum of ~1.1 eV to a maximum of ~4 eV). However, Pt decoration does favor the binding of both atomic (-2.17 eV) and molecular (-1.56 eV) oxygen, thereby keeping the possibility of adsorbed oxygen pathway active on the Zn-$t_1$ surface. Thus, we expect Pt decoration on ZnO to enable the adsorbed oxygen pathway on both the O-$t_1$ and Zn-$t_1$ terminations. Additionally, we expect water to bind reasonably well on both pristine (~-0.27 to –0.94 eV) and Pt decorated (~-1.17 to –0.25 eV) O-$t_1$ and Zn-$t_1$ surfaces, indicating that water can exist as a stable adsorbed species on both surfaces. $H_2O$ molecule binds best with the sub-surface Zn sites on the pristine O-$t_1$ surface (-0.42 eV), whereas it binds equally well at all sites on the pristine Zn-$t_1$ surface (-0.90 to -0.95 eV).

We evaluated the energies of various reaction pathways, using DFT, on both the O-$t_1$ and Zn-$t_1$ surfaces to evaluate the possible mechanisms that are active during device operation and compiled the results in **Figure 10**. Note that we did not consider transition or intermediate states of any reaction, focussing our efforts on the thermodynamic feasibility of possible reactions. We considered both $H_2O$ and OH formation as possible products, given that both species can cause a change in the carrier concentration thereby resulting in a sensor response. For calculating all reaction energies, we considered both reactants and products to be in their corresponding best binding sites (i.e., one of hollow, O, or Zn sites in pristine-ZnO or at Pt site for decorated ZnO).

In terms of OH formation, possible ways are through atomic O and H reactions and molecular $O_2$ and $H_2$ reactions (panels a and b in **Figure 10**, reaction energies in eV are normalised per OH formation), with both reactants and products being species that are adsorbed on one of the hollow, Zn, and O sites of pristine-ZnO or on the Pt site of a Pt decorated ZnO. Importantly, we find that OH formation is spontaneous (-1.50 eV) on the Pt decorated O-$t_1$ surface considering formation from atomic species, while pristine O-$t_1$ can't form OH from atomic O since the surface does not bind atomic O without Pt addition (see **Figure 9**a). OH formation is spontaneous on both pristine and Pt decorated O-$t_1$ (-1.26 to –1.43 eV) considering formation from molecular species. On the Zn-$t_1$ surface, OH formation is weakly favored (-0.08 to –0.21 eV) upon formation from atomic species while OH formation is strongly favored from molecular species (-1.32 to –1.38 eV). The formation of OH from molecular reactants are more favored likely due to O-H bonds being more stable than O-O or H-H bonds, which is also captured by the exothermic formation enthalpy of $H_2O$ in the gas state from $H_2$ and $O_2$.[52,53]

Thus, we expect OH formation to be broadly favored in both pristine and Pt decorated ZnO, with the Pt decoration making a difference with respect to pristine ZnO only on the O-$t_1$ surface.

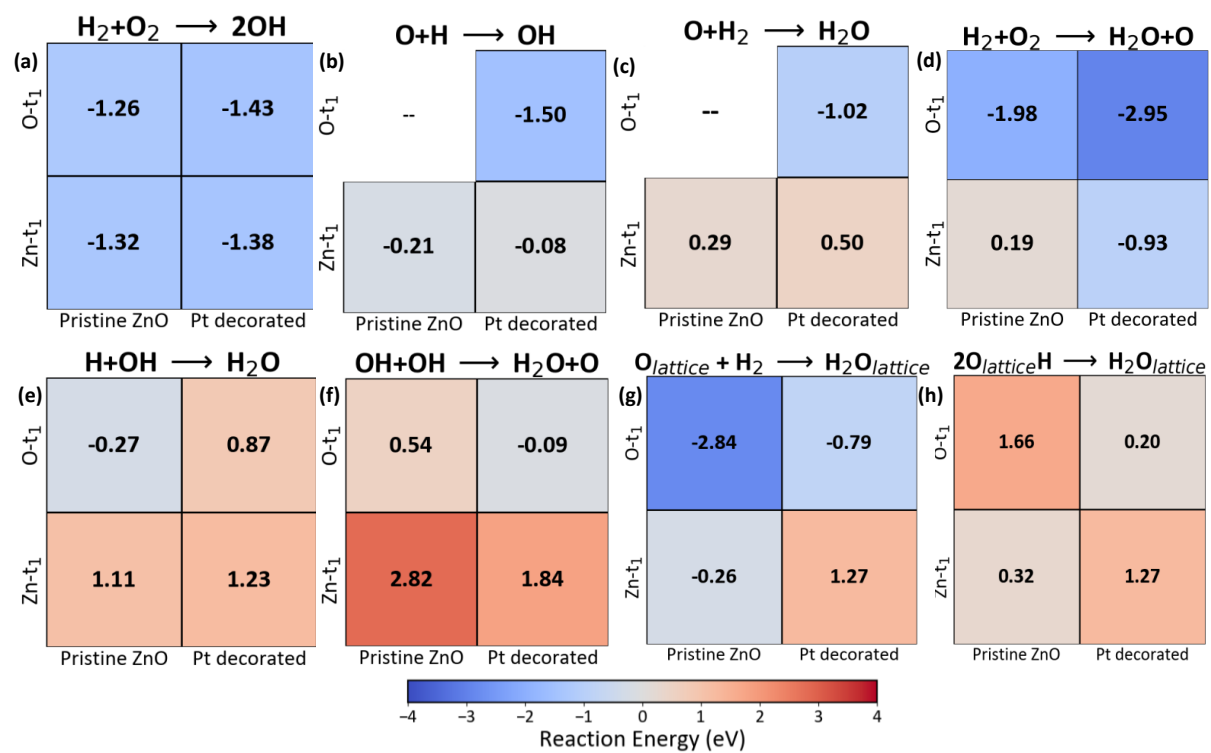

**Figure 10.** (a-b) OH formation energies from adsorbed oxygen and co-adsorbed hydrogen in both molecular and atomic states, respectively. Water formation energies via (c-d) adsorbed oxygen and co-adsorbed hydrogen pathway, (e-f) adsorbed OH with co-adsorbed H or OH pathway, and (g-h) lattice oxygen pathway.

We calculated the reaction energies to form water (all energies in eV normalised per H$_2$O molecule) based on four sets of adsorbed reactants, namely, atomic O and molecular H$_2$, molecular O$_2$ and H$_2$, atomic H and adsorbed OH, and two adsorbed OH (via an 'autoreduction' pathway), as displayed in panels c-f of **Figure 10**. Additionally, we considered water formation from the lattice oxygen (i.e., oxygen atoms that are intrinsic to the lattice and are not adsorbed species) via two mechanisms, namely using molecular adsorbed hydrogen and atomic adsorbed hydrogen (panels g and h, **Figure 10**). Given that water formation from lattice oxygens requires the creation of a lattice oxygen vacancy, i.e., the hydrogen removes an oxygen from the lattice to form water, we have compiled the oxygen vacancy formation energies on both O-$t_1$ and Zn-$t_1$ surfaces, with and without Pt decoration (**Figure S6**). Post Pt addition, the oxygen vacancy formation energy of the lattice oxygens near the cluster increases for both terminations. For all reactions, we considered the adsorbed O/O$_2$/OH to be at their best binding sites and the adsorbed H/H$_2$ to be on a neighboring best binding site. For reactions where the reactant is not a stable adsorbed species (e.g., atomic O on pristine O-t$_1$, see **Figure 9**) we have not computed reaction energies.

On the Zn-t$_1$ surface of pristine-ZnO, we observe that water formation is not thermodynamically favored via all reaction pathways, with the exception of adsorbed hydrogen molecule reacting with the lattice oxygen to form water (-0.26 eV, **Figure 10**g). With Pt decoration, water formation on Zn-t$_1$ is thermodynamically favored upon the reaction of molecular H$_2$ and molecular O$_2$ (-0.93 eV, **Figure 10**d), while the lattice oxygen pathway becomes unfavorable (1.27 eV, **Figure 10**g). Given that a reaction of molecular hydrogen with a lattice oxygen or co-adsorbed molecular oxygen may require the formation of atomic hydrogen (or OH) as an intermediate species, we expect the occurrence of these reactions to be low. Hence, we expect the pristine or Pt decorated Zn-t$_1$ surface to respond to hydrogen presence predominantly by forming adsorbed OH on the surface (panels a and b, **Figure 10**). However, the OH groups have strongly negative formation energies on the pristine Zn-t$_1$ surface and do not exhibit any favorable reaction energies towards water formation, indicating that the OH groups once formed will continue to remain on the Zn-t$_1$ surface. Such strong binding of OH groups may result in a drift in the response of the sensor (owing to occupation of available active sites) and affect its stability over multiple cycles, which we do not observe in our experiments with pristine-ZnO (**Figure 8d**). Thus, we expect the contributions of the Zn-t$_1$ surface to be minimal in the hydrogen sensor response observed with either pristine or Pt decorated ZnO.

In the case of the O-t$_1$ surface, three reaction pathways are possible to form in pristine ZnO, namely, atomic H and adsorbed OH (-0.27 eV, **Figure 10**e), molecular H$_2$ and O$_2$ (-1.98 eV, **Figure 10d**), and molecular hydrogen with lattice oxygen (-2.84 eV, **Figure 10g**). On the other hand, Pt decoration enables water formation via the adsorbed OH autoreduction pathway (-0.09 eV, **Figure 10f**), disables the reaction of adsorbed H and OH (0.87 eV, **Figure 10e**), and allows water formation via molecular H$_2$ with molecular O$_2$, and molecular H$_2$ with lattice O. Thus, the key difference between pristine and Pt decorated O-t$_1$ surface is the activation of the OH autoreduction pathway and deactivation of the adsorbed H+OH pathway. Given that the occurrence of reactions with molecular H$_2$ or O$_2$ species can be limited (as discussed above in the Zn-t$_1$ case), we hypothesize that the sensor response of the O-t$_1$ surface is caused predominantly by the H+OH pathway in the pristine state and the OH autoreduction pathway in the Pt decorated case, with the lattice oxygen pathway being active in both pristine and Pt decorated ZnO.

We can expect the availability of OH species to be higher in the Pt decorated O-$t_1$ compared to the pristine surface, since OH formation is favored through both adsorbed O + adsorbed H and molecular $O_2$ + molecular $H_2$ mechanisms in the Pt decorated surface (panels a and b of **Figure 10**). Thus, the autoreduction of a higher concentration of OH groups in the Pt decorated surface can result in a better sensor response compared to the H+OH mechanism that is active in the pristine surface. Therefore, we hypothesize that the O-$t_1$ surface termination of ZnO to predominantly contribute to our observed sensor response, with the adsorbed H+OH mechanism and lattice oxygen pathway being more active on the pristine, while Pt decoration boosts the sensor response by facilitating the autoreduction of OH groups that are adsorbed on the O-$t_1$ surface, alongside spontaneous adsorption and dissociation of molecular $H_2$ and keeping the lattice oxygen pathway active.

## 3. Discussion

The performance of our Pt decorated ZnO sensor demonstrates high competitiveness in terms of its low LOD for hydrogen, along with exceptionally fast response and recovery times, which is important in ensuring the viability of a hydrogen-dominated sustainable energy economy. To clearly showcase the superior sensing performance of our devices, we conducted a benchmarking analysis by comparing our sensors' results with those from similar studies reported in the literature (**Table S4**). We summarize our benchmarking in **Figure 11**, where we plot the sensor's response as a function of its hydrogen LOD (which provides a visual comparison of the sensitivity, panel a), and we compare the response and recovery times of our sensor in real-time detection compared to other studies (panel b). Our in-situ Pt decorated ZnO-based sensor achieves the lowest LOD, detecting hydrogen at concentrations as low as 0.1 ppm (100 ppb), with a 38% sensing response, as shown in **Figure 11**a. Moreover, our sensor exhibits significantly faster response and recovery times, just 10 sec and 3 sec, respectively, compared to similar gas sensors reported in the literature, as illustrated in **Figure 11**b. While several other sensors have been reported to detect ppb-level hydrogen, such sensors typically require higher operating temperatures compared to our Pt decorated ZnO sensors. Also, our Pt decorated sensor not only provides a low LOD for hydrogen but also a rapid response time compared to other devices, allowing it to detect low concentrations of hydrogen swiftly. Thus, we demonstrate the in-situ Pt decorated ZnO sensor as a highly competitive and efficient candidate for hydrogen gas sensing, standing out due to its low LOD, rapid response and recovery time, and stability at moderate temperatures compared to sensors developed previously.

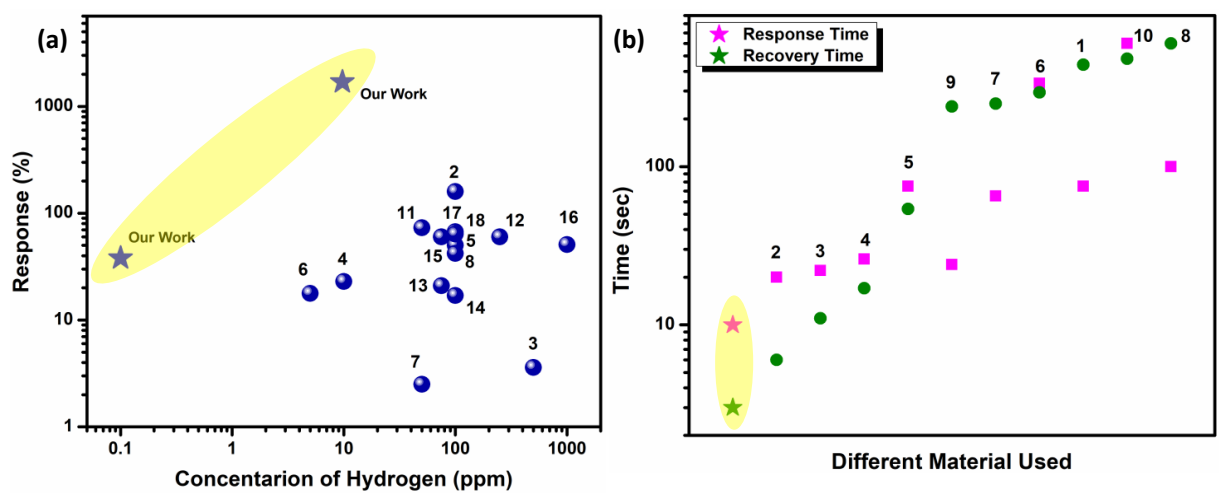

**Figure 11.** Comparison of our sensor with respect to literature data in terms of (a) response at different hydrogen concentrations and (b) response and recovery times. Yellow highlights or star symbols in both panels indicate values reported in our work.

Typically, semiconducting metal oxides interact with a variety of gases, lacking inherent selectivity.[6,54] However, the addition of platinum (Pt) to ZnO addresses these limitations by serving three important functions. First, Pt enhances the sensor's selectivity toward hydrogen ($H_2$). As illustrated in **Figure 7**d, cross-sensitivity tests demonstrate that while pristine ZnO sensors respond to multiple gases, Pt decorated ZnO exhibits a markedly higher response to hydrogen compared to other gases. This pronounced increase in hydrogen response ensures that hydrogen can be uniquely identified even in the presence of interfering gases. Second, Pt facilitates the efficient dissociation of $H_2$ molecules on its surface. Once $H_2$ dissociates at Pt sites, the resulting hydrogen radicals interact with surface-adsorbed oxygen or lattice oxygen species. This process, known as the spillover effect, is well-supported in the literature on noble metal decorated oxide substrates.[55] Third, Pt enables the autoreduction mechanism of OH groups on the ZnO surface, enabling them to form $H_2O$, which is normally easier to desorb than OH, thus increasing the stability of the sensor. Therefore, Pt not only enhances selectivity but also accelerates the reaction kinetics and enables essential reaction pathways for rapid sensor response and recovery.

Our reaction energy analysis of the lattice oxygen pathway, which we find to be active in both pristine and Pt decorated ZnO, aligns with earlier experimental observations on $SnO_2$.[49,50] Unlike previous studies which have focussed on the electronic structure changes with adsorption, our thermodynamic approach is in agreement with prior and current experimental data, emphasizing the often less-explored role of lattice oxygen in sensor operation. Thus, our finding supports the contributions of the lattice oxygen pathway, in addition to the typically cited adsorbed oxygen pathway. While our study provides valuable insights into the reaction mechanisms, it is important to acknowledge the limitations in our computational approach. First is the inherent trade-off between accuracy and computational feasibility. Our calculations employed the generalized gradient approximation (GGA) as implemented in the Perdew-Burke-Ernzerhof (PBE)[56] functional. While GGA provides a reasonable balance between accuracy and computational cost, more sophisticated functionals like the strongly constrained and appropriately normed (SCAN)[57] can yield improved accuracy.[58] Due to the significant computational expense associated with SCAN and the challenges in achieving convergence for our specific slab models, we opted for the computationally more tractable GGA. Future work may consider using SCAN or other meta-GGA functionals to improve the accuracy of binding and reaction energies.

We focused on the thermodynamics of reaction mechanisms at 0 K in this work, neglecting kinetic effects. Although we expect the identified mechanisms to remain qualitatively valid at higher temperatures, the relative rates and overall selectivity could be influenced by kinetic barriers. Additionally, we limited our calculations to the dominant (002) surface observed in XRD (**Figure 1**a) and we modelled the sputtered Pt using the smallest stable cluster, both of which are computational limitations of our study. Furthermore, we approximated the 'on Pt' site as the only unique adsorption site on the cluster, neglecting potential edge effects at the Pt-ZnO interface. A more thorough investigation of the Pt-ZnO interface and systematic exploration of various adsorption configurations, potentially using larger Pt clusters, can be useful as a follow-up work. Despite these limitations, we believe that our work provides valuable mechanistic insights, offering a foundation for future experimental and theoretical work on sensors.

## 4. Conclusion

We have developed a Pt decorated ZnO thin film sensor that delivers rapid response (10 sec) and recovery (3 sec) times at hydrogen concentrations of 10k ppm and detects hydrogen at levels as low as ~100 ppb. Utilizing an in-situ sputtering technique followed by Pt decoration, our sensor achieves stable and repeatable hydrogen detection from 100 ppb to 10 k ppm. We confirmed the stability through one year of repeated hydrogen exposure in a vacuum chamber, and validated the reproducibility with three independently fabricated samples. Our fabrication method, which involves surface decoration without any post-processing, is simple, scalable, and cost-effective. This approach not only enhances sensing performance but also ensures the production of stable thin films with a reproducible synthesis process. Consequently, our Pt decorated ZnO sensor is well-suited for large-scale production of efficient and reliable hydrogen sensing devices. Based on our DFT-based computations, we hypothesize that the O-exposed plane is the more active plane for sensing. The activation of multiple reaction pathways through Pt decoration accounts for the substantial improvements in sensor performance, including increased sensitivity, reduced operating temperature, and faster response times. Our findings highlight the pivotal role of noble metal decoration, using scalable process techniques, in enhancing the performance of metal oxide sensors for practical applications. Apart from the results, we would also like to emphasise the approach of this study, which systematically combines experimental and computational investigations to elucidate possible reaction mechanisms apart from demonstrating repeatable and robust sensor performance under various experimental conditions. Such approaches will help advance the development of the next generation of metal oxide-based sensors in a systematic and accelerated fashion, which we hope the community will take up in future studies.

## 5. Methods

### 5.1 Device fabrication

An IDE system with a uniform 5 μm spacing between electrodes was fabricated on a Si/SiO$_2$ substrate using a two-step process involving optical lithography and sputtering. The IDE pattern was defined through lithography, followed by the deposition of a Ti/Pt (10 nm/90 nm) metal layer via DC magnetron sputtering. A lift-off process was then used to form contact pads. The sensing material, a metal oxide thin film, was subsequently deposited onto the IDE's active region (1 mm² area) using radio frequency (RF) sputtering. The ZnO thin film was prepared through reactive RF magnetron sputtering from a high-purity (~99.99%) 3-inch ZnO target.

Sputtering was conducted in a chamber maintained at 6.8e$^{-3}$ mTorr, with a gas flow of 140 SCCM (standard cubic centimeters per minute) for chamber argon and 50 SCCM for magnetron argon, and the target-to-substrate distance was set at 7.5 cm. The chamber was evacuated to a base pressure of 3e$^{-6}$ mTorr prior to initiating the sputtering process, with pre-sputtering performed for approximately 1200 sec (900 sec with all gases and 300 sec with the final parameters). A 40 nm thick ZnO layer was then deposited on the IDE using a hard shadow mask. To enhance sensing performance, the top surface of ZnO was decorated with Pt through sputtering at varying deposition times (1, 2, 4, and 6 sec). The resulting sensor films were labelled as pristine ZnO, Pt_1s-ZnO, Pt_2s-ZnO, Pt_4s-ZnO, and Pt_6s-ZnO, respectively, corresponding to different Pt deposition times. The entire process flow is shown in **Schematic 1**. These sensor films were then subjected to further analysis and testing.

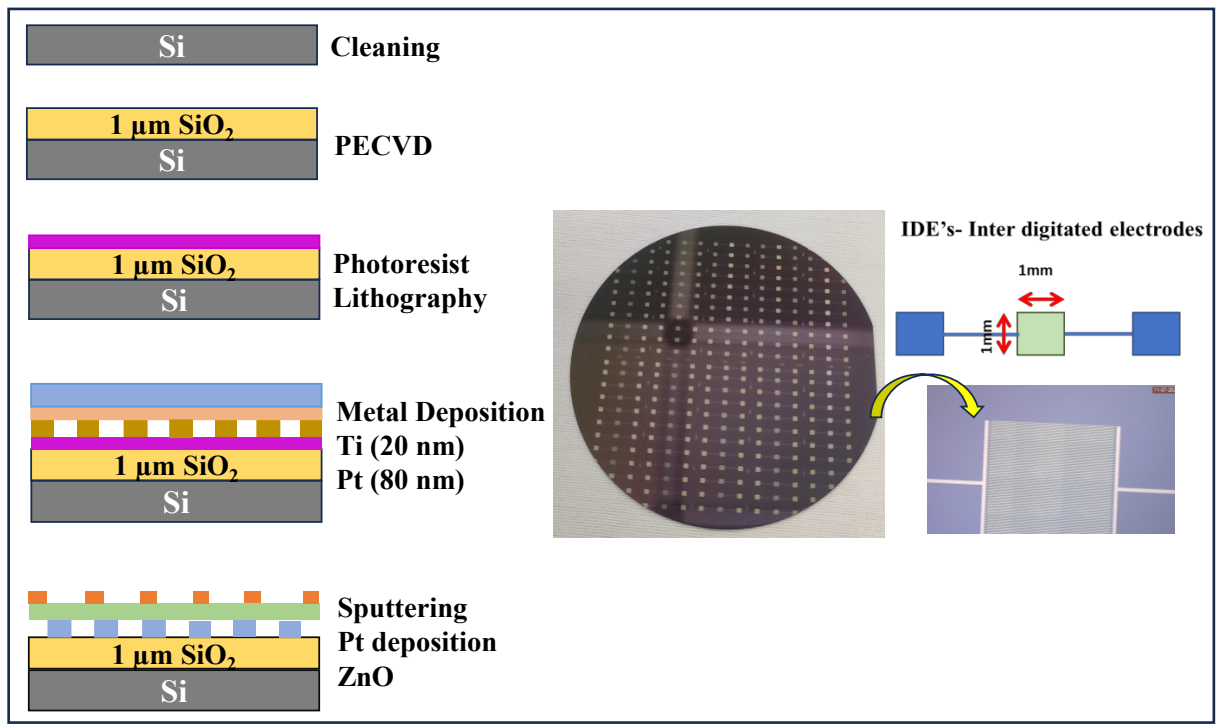

**Schematic 1.** Fabrication process flow of Pt decorated ZnO film

## 5.2 Computational Details

We calculated the bulk, surface, and adsorption energetics of polar surfaces using spin-polarized DFT [59,60] as implemented in the Vienna ab initio simulation package (VASP).[61,62] Projector augmented wave [63,64] (PAW) potentials were employed using a kinetic energy cutoff of 520 eV for the plane wave basis. We used a *k*-point density of 32 per Å to sample the reciprocal space (i.e., 32 sub-divisions sampled along a unit reciprocal lattice vector). To approximate the electronic exchange and correlation, we used the PBE functional.[65] We utilised

Grimme's zero damping DFT-D3[66] dispersion correction in all calculations. During structural optimization, we converged the residual forces between atoms to below |0.03| eV/Å and the total energies to within $10^{-5}$ eV/cell Dipole corrections were switched on during all relaxations. We have not explicitly corrected for the known oxygen molecule over binding with GGA [67]. However, adding a correction to the $O_2$ molecule will not change any qualitative trends or conclusions of the study.

We used slab models, where each slab extends is subject to periodic boundary conditions along the *a* and *b* axes, with the surface(s) lying perpendicular to the *c*-axis. We used 15 Å thick slabs with 15 Å of vacuum to separate periodic images along the *c*-axis. Slabs were created using the slab generator class of the pymatgen package.[68] We used selective dynamics during structure relaxation, i.e., only the top two layers along with any adsorbate were allowed to relax while all remaining layers were frozen to emulate bulk behaviour. Details on cluster geometry and pseudo hydrogen capping are provided in the SI.

**Supporting Information**

The supporting information file consists of additional details about experimental and computational section (S.I).

**Acknowledgments**

The authors acknowledge financial support from Shell India Markets Private Limited. We acknowledge the technical support provided by staff at Micro and Nano Characterization Facility (MNCF) at the Center for Nanoscience and Engineering (CeNSE) at the Indian Institute of Science (IISc), Bangalore, for the material characterization. The authors gratefully acknowledge the computational resources of the super computer 'PARAM Pravega' provided by Super Computer Education and Research Centre (SERC), IISc. The authors also acknowledge the computing resources provided to them on the high-performance computer noctua1 and noctua2 at NHR Centre PC2. This was funded by the Federal Ministry of Education and Research (www.nhr-verein.de/unsere-partner).

**Conflicts of interest**

The authors declare no competing financial interest.

**Data availability**

The computational data that support the findings of this study are openly available at our [GitHub](#) repository.

**Keywords**

Hydrogen, ZnO NPs, Pt decoration, IDE, Sputtering, Surface interaction, DFT, Material and Electrical Characterization, Oxygen Stoichiometry, Sensing mechanism.